\documentclass[a4paper]{jpconf}
\usepackage{graphicx}
\begin{document}
\title{ A construction of algebraic surfaces with many real nodes}
 \author{Juan Garc\'{\i}a Escudero}

\address{Universidad de Oviedo. Facultad de Ciencias Matem\'aticas y F\'{\i}sicas,
 33007 Oviedo, Spain}

\begin{abstract}
A family of algebraic surfaces with many nondegenerate real singularities is introduced with the help of a construction, which has been used in previous works for the generation of substitution tilings.
\bigskip\par
 { \it{Keywords}}: singularities of surfaces, real algebraic geometry.
  \bigskip\par

\end{abstract}

%

\section{Introduction}
\bigskip\par
In 1819 a quartic wave surface was discovered by Fresnel in the study of crystal optics. Later Kummer observed that the surface had the maximum possible number of nodes of a quartic in the complex projective space
${\bf{P}}^{3}({\bf{C}})$ and he constructed a surface with only real nodes. Since then the subject of algebraic surfaces has a rich history with plenty of achievements. 
\par
A construction of surfaces with degree $m$ in 
${\bf{P}}^{3}({\bf{C}})$ having only nondegenerate singular points, or double points, was introduced by Chmutov in \cite{chm92}. It was based on his previous idea of using Chebyshev polynomials and on the Givental methods for generating cubics \cite{arn85}.
\par
In \cite{bre08} it has been shown that the Chmutov construction can be adapted to give surfaces with only real singularities. In ${\bf{C}}^3$ a node or $A_{1}$ singularity can be written in the form $w_{1}^{2}+w_{2}^{2}+w_{3}^{2}=0$ in some local coordinates. In ${\bf{R}}^3$ there are two types of nodes, conical nodes ($x^{2}+y^{2}-z^{2}=0$) and solitary nodes ($x^{2}+y^{2}+z^{2}=0$) and the construction produces conical nodes. The main idea consists in defining real folding polynomials $F^{A_{2}}_{{\bf{R}}m}(x,y)\in {\bf{R}}[x,y] $ associated to the root lattice $\bf{A}_{2}$ (which are generalizations of Chebyshev polynomials, related with the root lattice $\bf{A}_{1}$).  In the cases with degrees 6,7,8,10,12, there are higher lower bounds for the maximum number of real nodes $\mu_{\bf{R}}$ on a surface of degree $m$. A 65-nodal sextic and a 345-nodal decic invariant under the symmetry group of the icosahedron were constructed by Barth \cite{bar96}. Other surfaces with the higher lower bounds known are Labs\'{}s 99-nodal septic, Endrass\'{}s 168-nodal octic, and the surface of degree 12 with $\mu_{\bf{R}}=600$ nodes, invariant under the reflection group of the regular four-dimensional 600-cell,  introduced by Sarti \cite{end97,lab05a,lab06,sar01}. The results in \cite{bre08} show that all known lower bounds for the maximum number of nodes on a surface of degree $m$, can be attained with only real singularities.
\par
In this article we consider a construction motivated by recent work on tilings generated by substitution or inflation rules. In particular, two subfamilies of surfaces with degree divisible by 3 are obtained. One is related to the dihedral group $D_{3}$ and the other to the cyclic group $C_{3}$. We use Mathematica \cite{wol91} and Surfer \cite{end01,lab05} computing and geometric visualization tools.  

 \begin{figure}[h]
 \includegraphics[width=22pc]{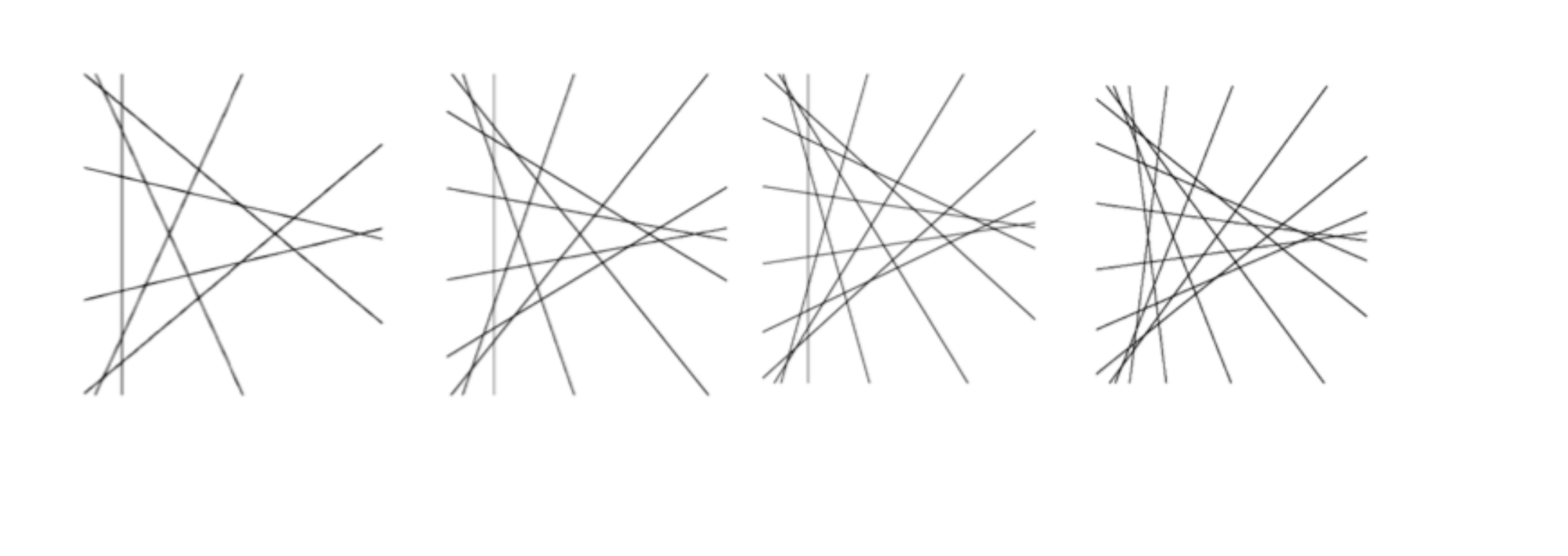}
\caption{\label{label} Simple arrangements $\Sigma_{D}=(D)_{odd}$ with seven, nine, eleven and twelve lines as subarrangements of the systems of lines (D). If $m=3q$ then $\Sigma_{D}$ has $D_{3}$ symmetry.}
\end{figure}

\section{The construction of surfaces with degree $m$}
\bigskip\par
 Arrangements of lines are labeled sets of lines not all passing through the same point. A vertex, which is the intersection of two or more lines, is called ordinary if only two lines meet. If all the vertices are ordinary, then the arrangement is said to be simple \cite{goo04}. 
 \par
 By a $d$-system we mean a simplicial arrangement of straight lines in $d$ orientations.  There are four types of geometric constructions, denoted (A), (B), (C) and (D) in \cite{esc08} 
which are the basis for the generation of tilings with arbitrarily high symmetry. They are of interest in the study of quasicrystals and other fields \cite{esc08, esc11} .
For our purposes we may consider just two types because the systems with $d$ odd are included in the systems with $d$ even: (B) is included as a subsystem of (A), and (D) is included in (C). 
In this work we denote the two basic constructions by (D) and (C) which correspond to (A) and (C) in \cite{esc08}.  We donote  $s_{\nu}\equiv sin (\nu \pi/d), t_{\nu}\equiv tan (\nu \pi/d)$. The following sets of lines in the $xy$ plane define, for each $d$, up to mirror images,  one or two finite patterns made with triangles having edge lengths $s_{\nu}$, and containing the necessary information for the derivation of substitution or inflation rules that generate aperiodic planar tilings:

  \begin{equation}
x=0, y=0, L_{\nu,d}^{X}(x,y)=0
\end{equation}

\par\noindent
where $L_{\nu,d}^{X}(x,y):=y-x t_{\nu}-\Gamma _{\alpha(\nu)}^{X}$ and the term $\Gamma _{\alpha(\nu)}^{X}$,  X=C,D is defined as follows
\bigskip\par
$\bullet$ (D) $d=2m$, ($m=3,4,...$)
 \par
For $\nu=1,2,...,m-1$, the index $\alpha(\nu)$ is defined as $\alpha(\nu)=\nu$  ($\nu=1,2,...,\lfloor m/2\rfloor$) , $ \alpha(\nu)=m- \nu$  ($\nu=\lfloor m/2\rfloor+1,...,m-1$) and $ \Gamma _{\alpha(\nu)}^{D}$  is defined by $ \Gamma _{\alpha(\nu)}^{D} := 
 - \sum_{k=1}^{\alpha(\nu)} s_{m+1-2 k }$. 
 For  $\nu=m+1,m+2,...,d-1$, we have $\alpha(\nu)=\nu-m$ ($\nu=m+1,...,m+ \lfloor m/2\rfloor$),  $ \alpha(\nu)=d- \nu$ ($\nu=m+ \lfloor m/2\rfloor+1,...,d-1$) and  $\Gamma _{\alpha(\nu)}^{D} :=  \sum_{k=1}^{\alpha(\nu)} s_{m+1-2 k }$.
\bigskip\par

$\bullet$ (C) $d=2m, m=3q$, ($q=1,2,...$)
  \par
For $\nu=1,2,...,m-2$ we have $ \alpha(\nu)=\nu$ ($\nu=1,2,...,m-3$), $ \alpha(\nu)=1$  if $\nu=m-2$ and
  $\Gamma _{\alpha(\nu)}^{C} := 
 -\sum_{k=1}^{\alpha(\nu)} s_{m-2 k}$.
   For $\nu=m-1$ the term $\Gamma _{\alpha(\nu)}^{C}$ is null and for $\nu=m+1,...,2m-1$ we have 
   $\Gamma _{\alpha(\nu)}^{C} := 
 \sum_{k=0}^{\alpha(\nu)} s_{m-2 k}$,  where in this case  
 $\alpha(\nu)=\nu-m$ ($\nu=m+1,...,m+\lfloor (m-1)/2\rfloor$) and   $\alpha(\nu)=2m-1-\nu$ ($\nu=m+\lfloor (m-1)/2\rfloor+1,...,2m-1$).
\bigskip\par
The $2m$-systems (X) are the union of two subsystems: $(X)_{even}$ contains $L_{\nu,2m}^{X}(x,y)=0$ for $\nu$ even and $(X)_{odd}$ has  $L_{\nu,2m}^{X}(x,y)=0$ for $\nu$ odd. We consider, in what follows, a kind of constructive duality between substitution tilings and surfaces with many singularities. The subsystems $(D)_{even},(C)_{odd}$ are on the basis of the general geometric constructions for substitution tilings with $m$-fold symmetry (see \cite{esc11} for the subsystems corresponding to $m=5,10$), which have been studied in \cite{esc08} and previous works \cite{esc99}. The subsystems $\Sigma_{D}=(D)_{odd}, \Sigma_{C}=(C)_{even}$ are simple arrangements that will be used as the starting point of the following construction of surfaces with degree $m$ having many real singularities. We define two families of polynomials associated with each simple arrangement:
 \begin{equation}
  \begin{array}{ll}
&J_{m,\Sigma_{D}}(x,y):=(-1)^{\lfloor \frac{m}{2}\rfloor} x^{\frac{1-(-1)^{m}}{2}} \prod_{\nu=1,2\nu-1\neq m}^{m}L_{2\nu-1,2m}^{D}(x,y)\in {\bf{R}}[x,y];  \\
&\\
&J_{m,\Sigma_{C}}(x,y):=(-1)^{\lfloor \frac{m+1}{2}\rfloor+1} x^{\frac{1+(-1)^{m}}{2}} y \prod_{\nu=1,2\nu \neq m}^{m-1}L_{2\nu,2m}^{C}(x,y)\in {\bf{R}}[x,y]
  \end{array}
\end{equation}

 \begin{figure}[h]
 \includegraphics[width=25pc]{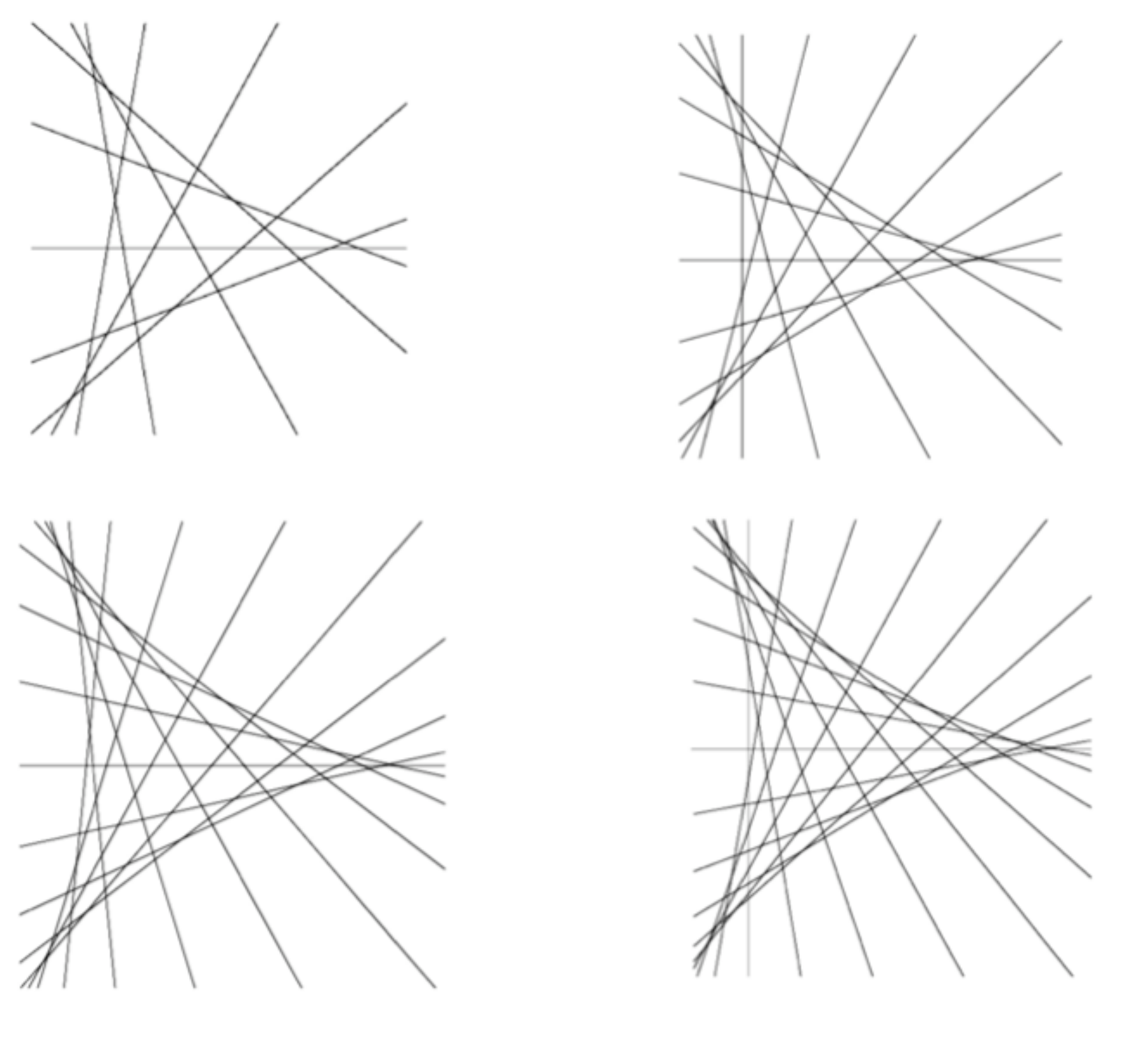}
\caption{\label{label} Simple arrangements $\Sigma_{C}=(C)_{even}$ with nine, twelve, fifteen and eighteen lines as subarrangements of the systems of lines (C). They have $C_{3}$ symmetry.}
\end{figure}

\par
Let $T_{m}(z)$ be the Chebyshev polynomials of the first kind generated by the recurrence relations $T_{m+1}(z)-2zT_{m}(z)+T_{m-1}(z)=0, m>0$ with $T_{0}(z)=1, T_{1}(z)=z $ (Fig.3). The real projective surfaces of degree $m$ associated with $\Sigma=\Sigma_{D},\Sigma_{C}$ are defined by 
  \begin{equation}
P_{m\Sigma}(x,y,z):=J_{m\Sigma}(x,y)+\frac{\lambda_{m\Sigma}}{2}( T_{m}(z)+1) \in {\bf{R}}[x,y,z]
\end{equation}
\par\noindent
where $-\lambda_{m\Sigma}$ is the unique critical value (minimum) of $J_{m\Sigma}(x,y)$ in the interior of the triangular regions of the simple arrangement $\Sigma$. In Fig.4, which corresponds to $m=9$, we have shown in black the regions where $J_{m\Sigma}(x,y)$ is negative and the minima are located inside the bounded triangular zones. The three different triangular shapes in $\Sigma_{D}$ and the seven triangular shapes in $\Sigma_{C}$ form two sets of prototiles for quasicrystal substitution tilings as shown in \cite{esc99}.
The polynomials $J_{m\Sigma}(x,y)$ have critical points with only three critical values 0, $-\lambda_{m\Sigma}$ and $\Lambda_{m\Sigma}$, therefore the surface $P_{m\Sigma}(x,y,z)=0$ is singular only at the points where the sum of the critical values of $J_{m\Sigma}(x,y)$ and $\frac{\lambda_{m\Sigma}}{2}( T_{m}(z)+1)$ is zero: either both are 0, or the first is $-\lambda_{m\Sigma}$ and the second $\lambda_{m\Sigma}$. The arrangements $\Sigma_{D}$ are equivalent to the simple arrangements corresponding to the real folding polynomials $F^{A_{2}}_{{\bf{R}}m}(x,y)$ in \cite{bre08}.  In fact it is possible to relate $J_{m\Sigma_{D}}(x,y)$ and $F^{A_{2}}_{{\bf{R}}m}(x,y)$ by introducing a change of variables and a scaling factor: 
$F^{A_{2}}_{{\bf{R}}m}(x,y)=b_{m}J_{m\Sigma_{D}}(a_{m}x+a_{m},a_{m}y)$ with $b_{m}=\frac{2}{a_{m}^m}, a_{m} =\frac{sin \frac{(m+2)\pi}{4m} sin \frac{(m-2)\pi}{4m}sin \frac{ \pi}{m}}{sin \frac{ \pi}{2m}sin \frac{2 \pi}{m}}$ if $m$ is even and $b_{m}=\frac{2m}{a_{m}^m}, a_{m} =\frac{2 cos \frac{\pi}{2m} sin^{2} \frac{ \pi}{m}}{3 sin \frac{\pi}{m}-sin \frac{3 \pi}{m}}$ if $m$ is odd. In Fig.1 we have shown the cases $m=7,9,11,12$ for $\Sigma_{D}$. When $m=3q$, the symmetry group of $\Sigma_{D}$ is the dihedral group $D_{3}$. 
\par
The surfaces $P_{m\Sigma_{D}}(x,y,z)=0$ and  \cite{bre08} 
  \begin{equation}
Chm^{A_{2}}_{{\bf{R}}m}(x,y,z):=F^{A_{2}}_{{\bf{R}}m}(x,y)+\frac{1}{2}( T_{m}(z)+1)=0
\end{equation}
\par\noindent
have the same number of nodes. The number of points of $J_{m\Sigma}(x,y)$ with critical value zero is equal to the number of vertices in the simple arrangements, namely $m \choose 2$. The number of triangles in $\Sigma_{D}$ is $\frac{1}{3}m^{2}-m$ if $m=0$ mod 3 and $\frac{1}{3}m^{2}-m+\frac{2}{3}$ otherwise. On the other hand $T_{m}(z)$ have $\lfloor \frac{m}{2}\rfloor$ points with critical value $-1$ and $\lfloor \frac{m-1}{2}\rfloor$ points with critical value $1$, hence the number of real nodes of $P_{m\Sigma_{D}}(x,y,z)=0$ and $Chm^{A_{2}}_{{\bf{R}}m}(x,y,z)=0$ is $\mu_{\bf{R}}={m \choose 2}\lfloor \frac{m}{2}\rfloor+(\frac{1}{3}m^{2}-m)\lfloor \frac{m-1}{2}\rfloor$ if $m \in {\bf{Z}}_{3}$ and $\mu_{\bf{R}}={m \choose 2}\lfloor \frac{m}{2}\rfloor+(\frac{1}{3}m^{2}-m+\frac{2}{3})\lfloor \frac{m-1}{2}\rfloor$ otherwise. 

\begin{figure}[h]
\includegraphics[width=15pc]{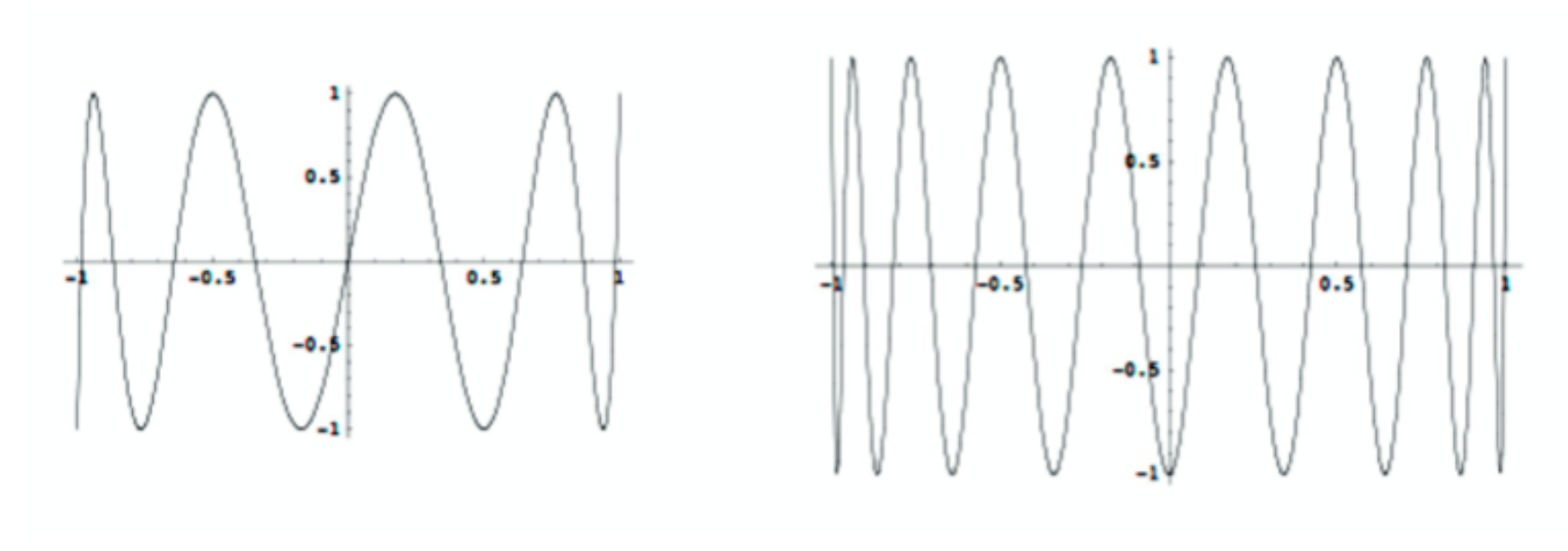}
\caption{\label{label}The Chebyshev polynomials $T_{9}(z)$ and $T_{18}(z)$.}
\end{figure}

 \begin{figure}[h]
 \includegraphics[width=15pc]{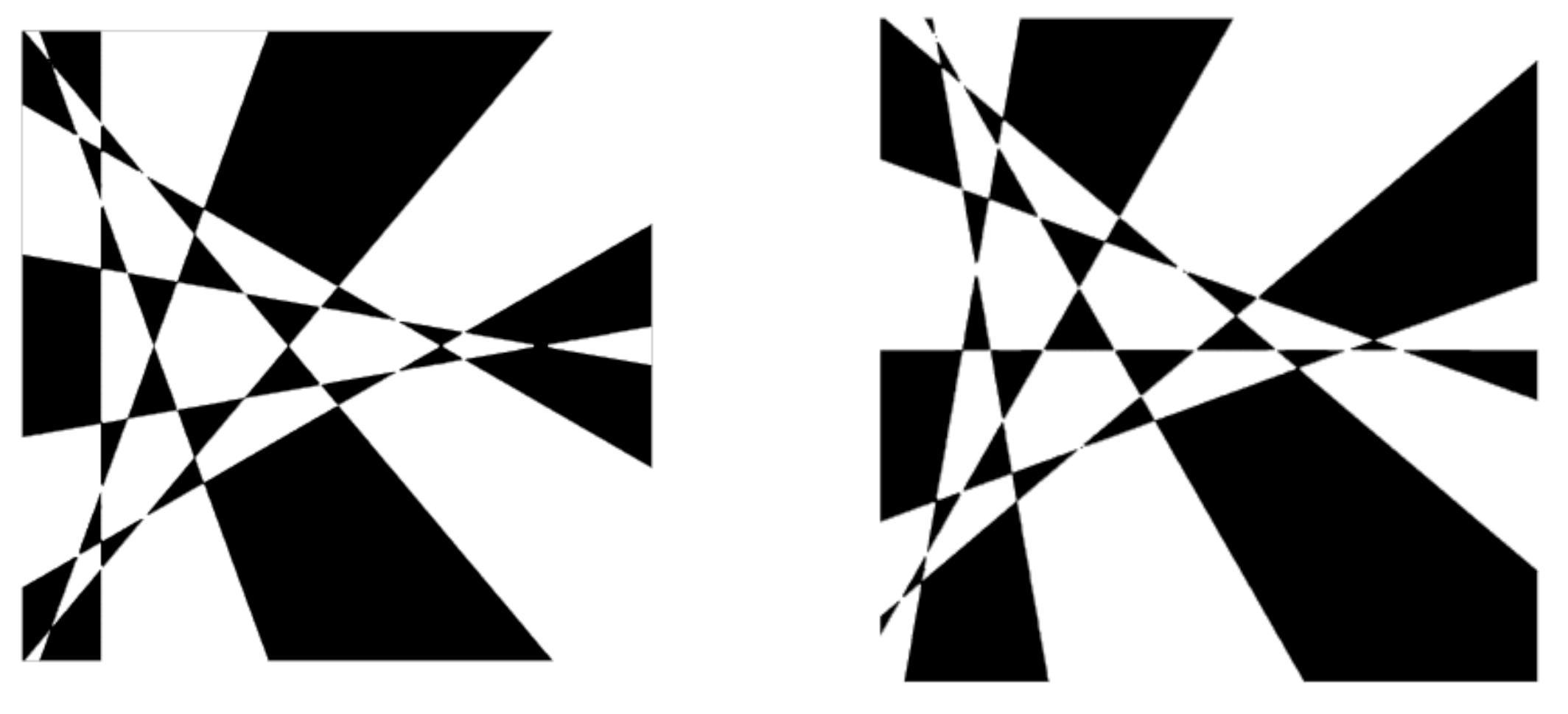}
\caption{\label{label}The polynomials $J_{9\Sigma_{D}}(x,y)$ (left) and $J_{9\Sigma_{C}}(x,y)$ (right) have negative values within the regions in black, and minima inside the bounded black triangles.}
\end{figure}

\section{The surfaces with degree $m=3q$ associated with $\Sigma_{C}$}
\bigskip\par
Within $\Sigma_{C}$ there is one more triangle than in $\Sigma_{D}$ while the number of vertices is the same, therefore the corresponding surfaces for $m=3q$ have more singularities. As we have done in the previous section, in relation with $\Sigma_{D}$, we can also make a change of variables in $J_{m\Sigma_{C}}(x,y)$, which facilitates our analysis. We translate the origin to the barycenter of the unique equilateral triangle in $\Sigma_{C}$, having coordinates $(c_{3q},\frac{1}{4})$ with $c_{3q}=\frac{\sqrt{3}}{4}+\frac{ sin((q-1) \pi/6q)}{2 sin(\pi/6q)}$. The parameter corresponding to the minimum in eq.(3) is then 
   \begin{equation}
 \lambda_{3q\Sigma_{C}}=J_{3q\Sigma_{C}}(c_{3q},\frac{1}{4})
   \end{equation} 
  The symmetry group of the simple arrangements $\Sigma_{C}$ is the cyclic group $C_{3}$ (Fig.2), hence for the study of the critical points of $J_{m\Sigma_{C}}(x,y)$ is enough to consider the $\frac{m(m-3)}{9}+1$ different triangular shapes which are, on the other hand, the prototiles of certain classes of substitution tilings \cite{esc99,esc08}. The action of $C_{3}$ gives an orbit of length 3 except for the central triangle therefore we have $\frac{1}{3}m^{2}-m+1$ minima. A first set of minima (which correspond to the triangles above the horizontal line and with an edge on it in Fig.2) can be obtained by rotating $\Sigma_{C}$ an angle of $\Theta_{m}=\frac{\pi}{6m}$ and the remaining ones with rotations by integer multiples of $\Theta_{m}$. We also change the scale in order to have all the local minima equal to $-1$ and all the local maxima equal to $8$. We define for $m=3q$
\begin{equation}
P_{m}^{C}(x,y,z):=J_{m}^{C}(x,y)+\frac{1}{2}(T_{m}(z)+1) \in {\bf{R}}[x,y,z]
\end{equation}
\par\noindent
with
  \begin{equation}
  \begin{array}{lll}
  J_{m}^{C}(x,y):=\frac{1}{\lambda_{m\Sigma_{C}}}J_{m\Sigma_{C}}(u,v)&&\\
 &&\\ 
  u=a_{m} x cos\Theta_{m} + a_{m} y sin\Theta_{m} +c_{m}, &v=-a_{m} x sin\Theta_{m} + a_{m} y cos\Theta_{m} +\frac{1}{4}&\\ 
   &&\\ 
  \end{array}
\end{equation}
 \par\noindent
where the coefficients $a_{m}$ have been given in Section 2.
 
 \begin{figure}
\includegraphics[width=22pc]{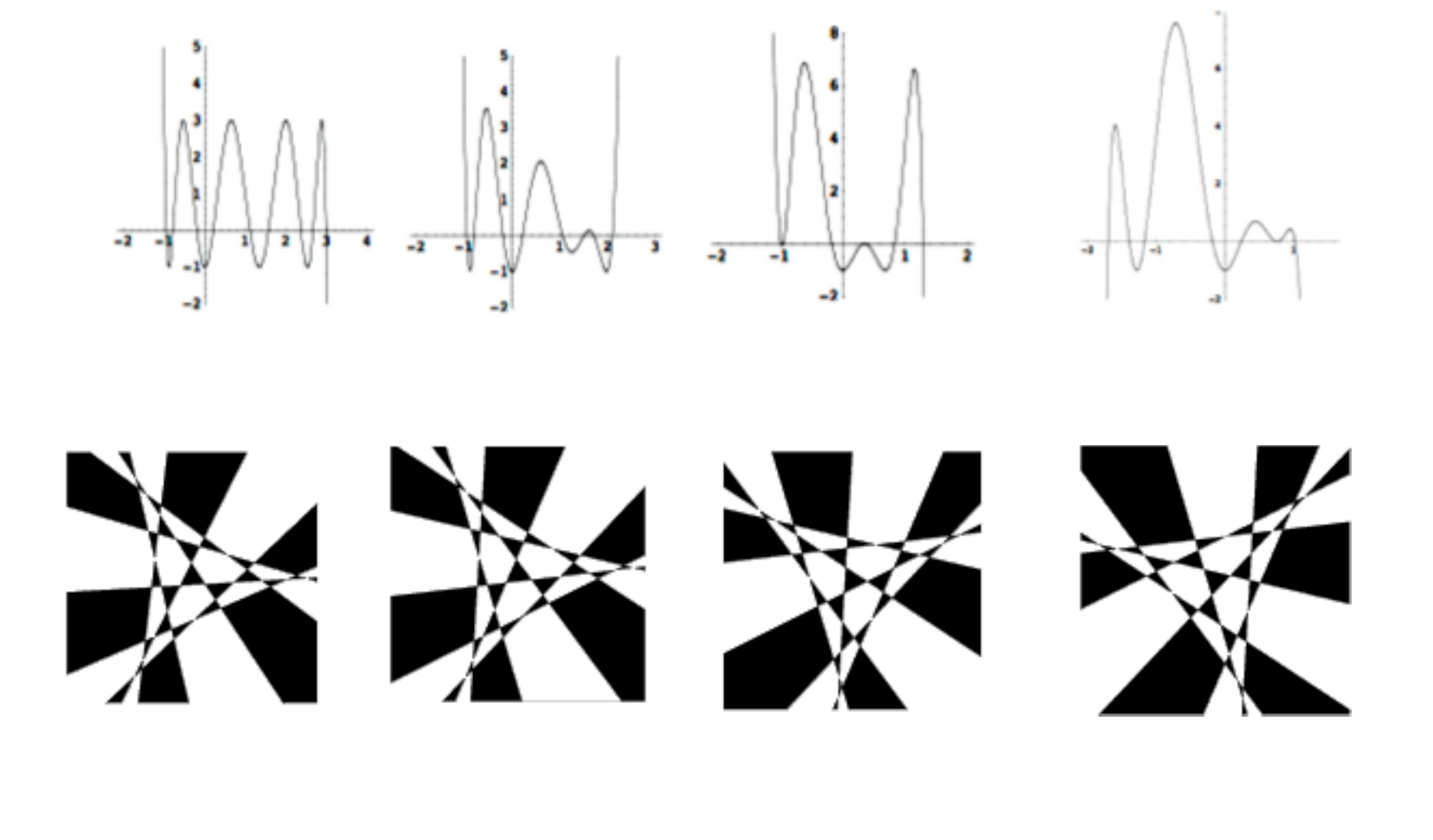}
  \caption{Local minima with value -1 obtained when  $\Sigma_{C}$ for $m=9$ is rotated by $k \pi/54$. From left to right: $\pi/54$ (prototiles $f,g,e,d$),$2\pi/54 (g,a) $, $8\pi/54 (g,c),14\pi/54 (b,g)$. } 
\end{figure}

 \begin{figure}
\includegraphics[width=22pc]{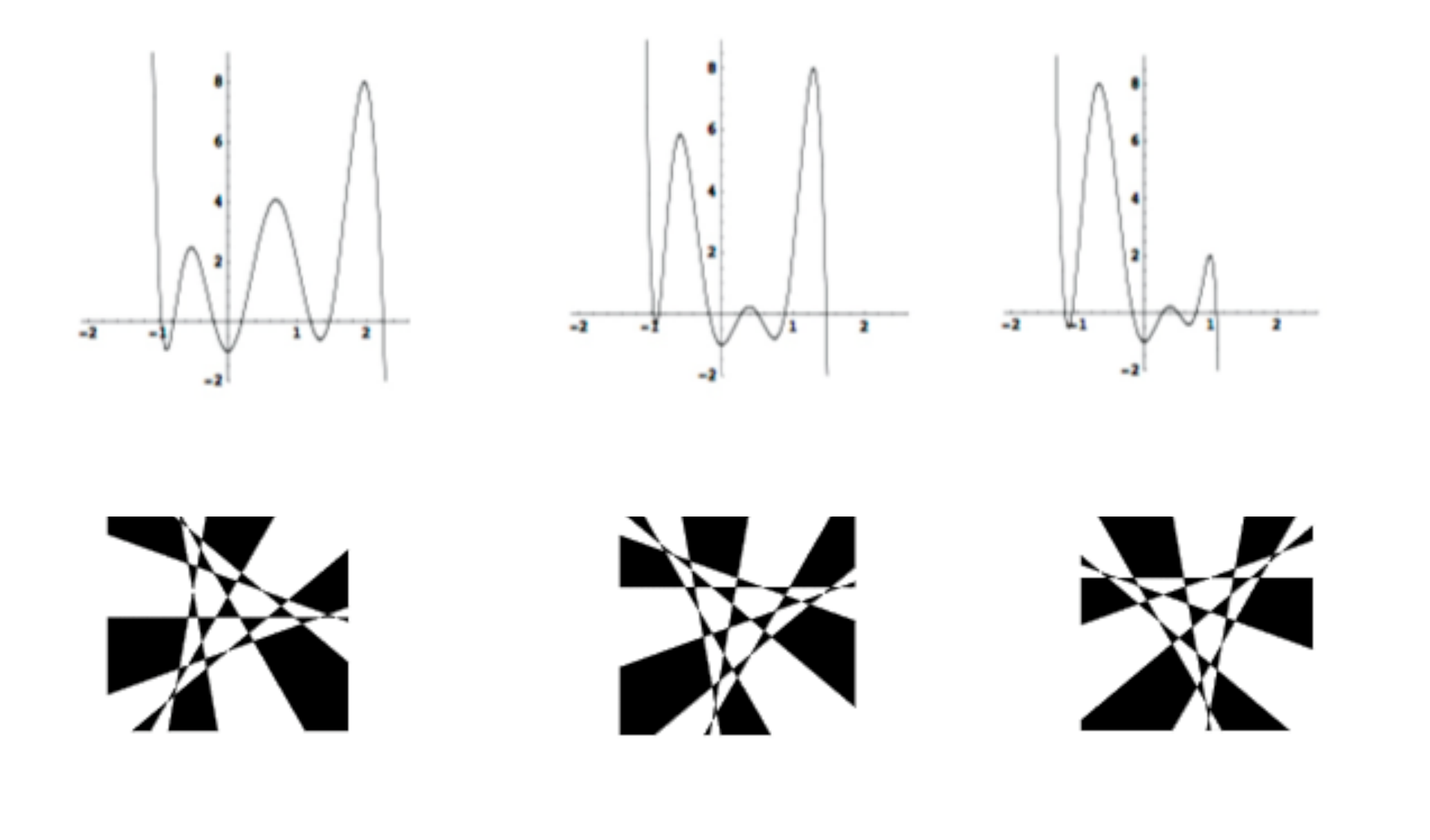}
  \caption{Local maxima with value 8 obtained when $\Sigma_{C}$, for $m=9$, is rotated by $k \pi/54$ with $k=0,6,12$ (from left to right).} 
\end{figure}
 
 \par
   The surface of degree 9 corresponding to $\Sigma_{C}$ in eq(3), contains the polynomial
  \begin{equation}
 \begin{array}{lll}
J_{9\Sigma_{C}}(x,y)&=&(y -x t_{2} +  s_{7} +s_{5} )
(y - x t_{4}+s_{7} +s_{5} + s_{3} +s_{1})
 (y - x t_{6}+ s_{7} + s_{5})
 (y - xt_{8})\\
   &&
 (y - xt_{10} - s_{9}- s_{7})
 (y - xt_{12}-s_{9} - s_{7}- s_{5} - s_{3})
  (y - xt_{14} - s_{9}- s_{7} - s_{5} - s_{3})\\
   &&
  (y - xt_{16} - s_{9} - s_{7})y
      \end{array}
\end{equation}
 \par\noindent
and $\lambda_{9\Sigma_{C}}=\frac{c_{9}^{9}}{2}\approx 13.28$. 
By using the notation in \cite{esc99}  we have three scalene triangular prototiles $a,b,c$ in the 9-system $(D)_{even}$ and in $(C)_{odd}$ we have four additional triangles: one is equilateral ($g$) and three are isosceles ($d,e,f$). In $\Sigma_{C}$ (Fig.4) the triangles above the horizontal axis and with an edge on it are, from left to right  $f,g,e,d$. The tile $b$ has a common vertex with $f$ and $c$ shares a vertex with $g$ and another with $e$. In this case $J_{9}^{C}(x,0)=-1+27x^{2}-9x^{3}-54x^{4}+36x^{5}+21x^{6}-27x^{7}+9x^{8}-x^{9}$ has four minima corresponding to $f,g,e,d$. The minima inside $a,c,b$  are obtained by restriction to the x-axis of the polynomials obtained from eq.(7) when $\Theta_{m}$ is replaced by the rotation angles $2\pi/54, 8\pi/54, 14\pi/54$, respectively (the minimum in the negative x-axis for $2\pi/54$ in Fig.5 is not -1 and does not correspond to a local extremum inside a prototile). The whole set of minima can be obtained with the rotation angles $\pi/54+l \pi/3,(2+6k)\pi/54+l\pi/3; k,l=0,1,2$ (Fig.5).  The maxima are located in nine bounded non-triangular regions with three different shapes. They can also be reached by rotations of $k\pi/54$ (see Fig.6 for the cases with $l=0$): $6k\pi/54+l\pi/3; k,l=0,1,2$. There are ${9 \choose 2}=36$ vertices where $J_{9\Sigma_{C}}(x,y)$ has critical value 0 and the minima are located inside each of the 19 black triangles in Fig.4. Now $T_{9}(z)$ has critical value -1 in four points and 1 in also four points, therefore the number of nodes of $P_{9}^{C}(x,y,z)=0$ is $\mu_{\bf{R}}=220$.
\par
The polynomial $J_{15\Sigma_{C}}(x,y)$ has the same minimum value $-\lambda_{15\Sigma_{C}}=-\frac{c_{15}^{15}}{2}\approx -2.4$x$10^{5}$ inside the 61 triangles of $\Sigma_{C}$ (Fig.2). Now $J_{15}^{C}(x,0)=-1+75x^{2}-25x^{3}-450x^{4}+300x^{5}+895x^{6}-945x^{7}-495x^{8}+1045x^{9}-297x^{10}-285x^{11}+260x^{12}-90x^{13}+15x^{14}-x^{15}$ has seven minima with value -1, and the minima of the other 14 prototiles are located by rotations of type $\frac{k \pi}{90}, k \in {\bf{Z}}$. The surface $P_{15}^{C}(x,y,z)=0$ has $\mu_{\bf{R}}=1162$. 
\par
For $m$ even $\lambda_{m\Sigma_{C}}=\frac{a_{m}^{m}}{2m}$ and $\lambda_{18\Sigma_{C}}\approx 4.8$x$10^{6}$. The arrangement $\Sigma_{C}$ has now 91 triangular regions (Fig.2), the polynomial $J_{18}^{C}(x,0)=-1+108x^{2}-36x^{3}-945x^{4}+630x^{5}+2919x^{6}-3024x^{7}-3366x^{8}+5720x^{9}-4212x^{11}+2457x^{12}+378x^{13}-1035x^{14}+528x^{15}-135x^{16}+18x^{17}-x^{18}$ has 8 minima, the directions for finding the other minima are given by $\frac{k \pi}{108}, k \in {\bf{Z}}$, and the number of nodes of $P_{18}^{C}(x,y,z)=0$ is $\mu_{\bf{R}}=2105$. 
 
\section{ The local extrema of $J_{m}^{C}(x,y)$ and the number of real nodes of $P_{m}^{C}(x,y,z)=0$.}
\bigskip\par
It is possible to get an alternative formulation which allows to obtain the positions of the extrema, having in mind that they are situated in the directions given by $\frac{k \pi}{6m}$. For $m=3q, q=1,2,3,...$ we define the lines $\bar{L}_{k,m}(x,y)=0$ with
     \begin{equation}
\bar{L}_{k,m}(x,y):=y-(cos \frac{k 2 \pi}{6m}-x)tan\frac{k \pi}{6m}-sin \frac{k 2 \pi}{6m}
 \end{equation}   
 where $k\in \bf{Z}$ and $\bar{L}_{3m,m}(x,y)=0$ is interpreted as the line $x=-1$. The polynomials obtained with $\bar{L}_{k,m}(x,y)$ are defined as 
      \begin{equation}
\bar{J}_{m}^{C}(x,y):= 3^{\frac{1-(-1)^{m}}{4}} (-1)^{\lfloor \frac{q+1}{2}\rfloor+1} \prod_{\nu=0}^{m-1}\bar{L}_{6\nu+1,m}(x,y)\in {\bf{R}}[x,y]
 \end{equation} 
  \par
  The non-integer coefficients $\alpha$ in $J_{m}^{C}(x,y)$ and $\bar{J}_{m}^{C}(x,y)=J_{m}^{C}(x,-y)$ appear only in the monomials $\alpha x^{n}y^{k}$ with $k$ odd. On the other hand, all the monomials in $F^{A_{2}}_{{\bf{R}}m}(x,y)$ have integer coefficients and, in addition to the relation given in Section 2, we have another one for the folding polynomials:  
  $F^{A_{2}}_{{\bf{R}}m}(x,y)=6-J_{m}^{C}(x,y)-\bar{J}_{m}^{C}(x,y)$.
   \par
 In order to locate the minima we consider the sets of lines $M_{-1}=\{\bar{L}_{6\nu,m}=0,\bar{L}_{6\nu+2,m}=0\}, \nu=0,1,...,m-1$. The minima are placed at all the points where exactly three lines of $M_{-1}$ meet (Fig.7).  We relabel the lines as $l_{k}, k=1,2,...2m$, with $l_{1}: \bar{L}_{0,m}=0$ and the others are assigned as they appear in consecutive clockwise orientations. When $\{l_{k1}\cap l_{k2}\cap l_{k3}\}$ is non-empty, then the subindices are either all even with $k1+k2+k3=2m+4$, or all odd  with $k1+k2+k3=2m+3$, modulo $2m$. The set of minima in the $x$-axis are (we indicate only two of the three intersecting lines): 
    \begin{equation}
 \cup_{k=0}^{\lfloor \frac{m-3}{2}\rfloor} \{l_{1}\cap l_{3+2k}\} 
  \end{equation} 
  \par\noindent
  The minimum in the central equilateral triangle corresponds to $k=q-1$ and the number of minima in the $x$-axis is $1+\lfloor\frac{m-3}{2}\rfloor$. The remaining minima, connected with intersections with odd subindices, are located, up to 3-fold rotation and whenever the subindices are positive, in 
      \begin{equation}
  \begin{array}{ll}
\{l_{3}\cap l_{5},l_{3}\cap l_{7},...,l_{3}\cap l_{2q-3}\} \cup\{l_{6q-1}\cap l_{6q-3},l_{6q-1}\cap l_{6q-5},...,l_{6q-1}\cap l_{6q-(2q-5)}\} &\\
\{l_{5}\cap l_{7},l_{5}\cap l_{9},...,l_{5}\cap l_{2q-5}\} \cup\{l_{6q-3}\cap l_{6q-5},l_{6q-3}\cap l_{6q-7},...,l_{6q-3}\cap l_{6q-(2q-7)}\}& \\
\{l_{7}\cap l_{9},l_{7}\cap l_{11},...,l_{7}\cap l_{2q-7}\} \cup\{l_{6q-5}\cap l_{6q-7},l_{6q-5}\cap l_{6q-9},...,l_{6q-5}\cap l_{6q-(2q-9)}\}& \\
...
   \end{array}
  \end{equation} 
    \par\noindent
  The last term of the sequence is given by $\{l_{q-1}\cap l_{q+1}\}$ if $q$ is even, or $\{l_{q-2}\cap l_{q},l_{q-2}\cap l_{q+2}\}$ if $q$ is odd. Now we have $2\sum_{k=0}^{\frac{q}{2}-2}(2k+1)$ minima if $q$ is even and $2\sum_{k=0}^{\frac{q-1}{2}-2}(2k+2)$ if $q$ is odd. Minima appearing in the intersections with even subindices are located in 
          \begin{equation}
  \begin{array}{ll}
\{l_{2}\cap l_{4},l_{2}\cap l_{6},l_{2}\cap l_{8},...,l_{2}\cap l_{2q}\} \cup\{l_{4}\cap l_{6},l_{4}\cap l_{8},...,l_{4}\cap l_{2q}\}\cup &\\
\cup\{l_{6}\cap l_{8},l_{6}\cap l_{10},...,l_{6}\cap l_{2q}\} \cup...\cup\{l_{2q-2}\cap l_{2q}\}& \\
   \end{array}
  \end{equation} 
      \par\noindent
      which gives $\sum_{k=0}^{q-1}k$ points. Due to the cyclic symmetry, if a minimum appears in $l_{k1}\cap l_{k2}$ then there are two more situated in $l_{k1+2qn}\cap l_{k2+2qn}\cap l_{2m+3-k1-k2+2qn}, n=1,2$. The total amount of critical points with value $-1$ is then $1+\frac{m(m-3)}{3}$. 

   \par
  Also the maxima are located at all the points of intersections of three lines ( Fig.7 ), belonging now to the set $M_{8}=\{\bar{L}_{6\nu+4,m}=0\}, \nu=0,1,...,m-1$ which is a $m$-system of type $(C)_{odd}$. In this case the lines are $l_{k}, k=1,2,...m$, with $l_{1}: \bar{L}_{4,m}=0$.  For those sets $\{l_{k1},l_{k2},l_{k3}\}$ intersecting in one point we have that their subindices satisfy $k1+k2+k3=1$, modulo $m$.
  The maxima are found, up to 3-fold rotation, in the following $\sum_{k=0}^{q-1}k$ points
          \begin{equation}
  \begin{array}{ll}
\{l_{1}\cap l_{2},l_{1}\cap l_{3},l_{1}\cap l_{4},...,l_{1}\cap l_{q}\} \cup\{l_{2}\cap l_{3},l_{2}\cap l_{4},...,l_{2}\cap l_{q}\}\cup &\\
...\cup\{l_{q-2}\cap l_{q-1},l_{q-2}\cap l_{q}\}\cup\{l_{q-1}\cap l_{q}\}& \\
   \end{array}
  \end{equation} 
      \par\noindent
The remaining ones are located in $l_{k1+qn}\cap l_{k2+qn}\cap l_{1-k1-k2+qn}, n=1,2$, when $l_{k1}\cap l_{k2}$ is included in eq(14). Adding up all the points we get $\frac{m(m-3)}{6}$ maxima with value $8$.
 \begin{figure}
 \includegraphics[width=38pc]{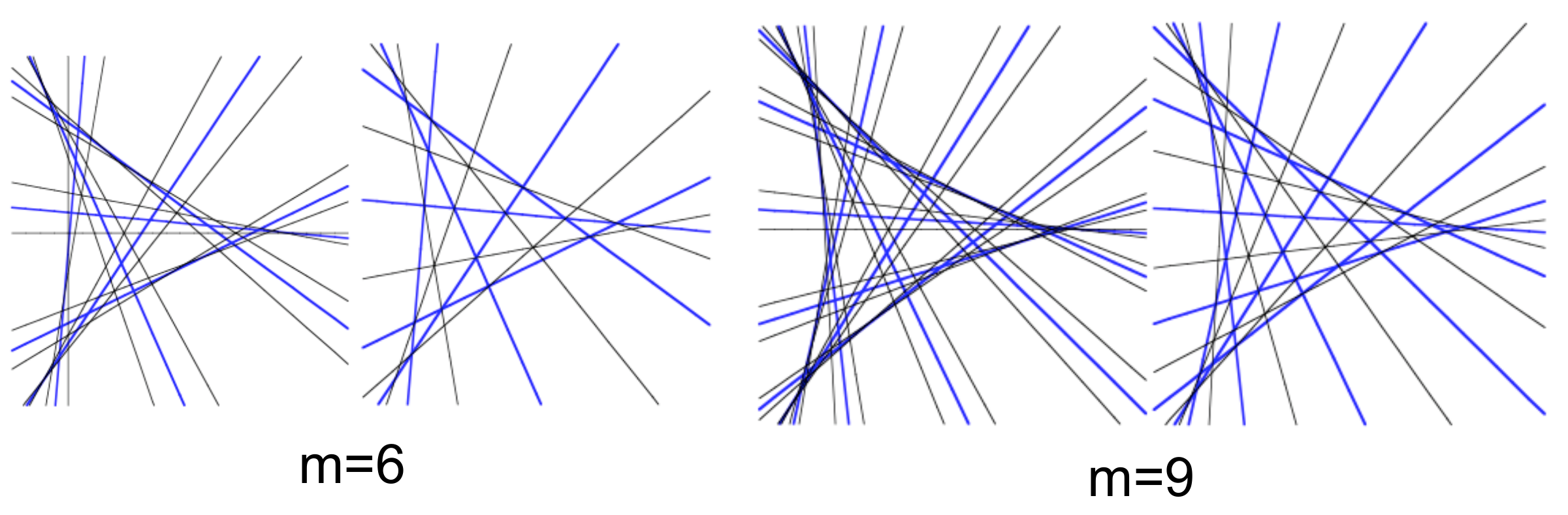}
  \caption{Local minima and maxima of $\bar{J}_{m}^{C}(x,y)$ are placed at all the points where three black lines, belonging to the sets $M_{-1}$ and $M_{8}$ respectively,  meet. The simple arrangements are represented in blue.}
\end{figure}

\par
An alternative way of obtaining the number of minima consists in observing that $M_{-1}$ is formed by the union of the $m$-systems $(D)_{even}$ and $(C)_{odd}$ (a mirror reflection of the one appearing in the analysis of the maxima). There are $1+\frac{m(m-3)}{6}$ points of intersection of three lines in $(D)_{even}$ and $\frac{m(m-3)}{6}$ in $(C)_{odd}$.  By taking into account the total number of minima we can compute the number of real nodes of $P_{m}^{C}(x,y,z)=0$ and the result is
  \begin{equation}
\mu_{\bf{R}}={m \choose 2}\lfloor \frac{m}{2}\rfloor+(1+\frac{m(m-3)}{3})\lfloor \frac{m-1}{2}\rfloor
\end{equation}
 \par\noindent
which exceeds in $\lfloor \frac{m-1}{2}\rfloor$ the number of nodes of $P_{m\Sigma_{D}}(x,y,z)=0$ or $Chm^{A_{2}}_{{\bf{R}}m}(x,y,z)=0$ for $m=3q$. Instead of the polynomials $T(z)$, we can use $J_{m}^{C}(z,0)$ and we obtain surfaces with the same number of nodes: 
  \begin{equation}
Q_{m}^{C}(x,y,z):=J_{m}^{C}(x,y)+\frac{(-1)^{m+1}}{4}(J_{m}^{C}(z,0)-1+(-1)^{m+1}2)\in {\bf{R}}[x,y,z]
\end{equation}
 \par\noindent
 The surfaces  $Q_{m}^{C}(x,y,z)=0, m=6,9,12,15,18$, visualized with the Surfer program, which shows the images within an invisible sphere, are represented in Fig.8. They have mirror partners  $\bar{Q}_{m}^{C}(x,y,z)=0$ defined as in eq(16) by replacing $J_{m}^{C}$ with  $\bar{J}_{m}^{C}$ ( $\bar{Q}_{6}^{C}(x,y,z)=0$ can be seen in Fig.8). We observe that this is not the case for the surfaces $P_{m\Sigma_{D}}(x,y,z)=0$ or $Chm^{A_{2}}_{{\bf{R}}m}(x,y,z)=0$, because $\Sigma_{D}$ has one ($m\neq 3q$) or three ($m=3q$)  mirror symmetry planes (Fig.1).
\par 
The polynomials introduced in this work can also be used for the construction of hypersurfaces with many real nodes. For instance, in addtion to the Chmutov hypersurfaces in 4-dimensional space defined with the folding polynomials \cite{chm92} we have, in the cases corresponding to $m=3q$, new types of pairs of mirror three-folds with affine equations  $J_{m}(x_{0},x_{1},x_{2},x_{3})=J_{m}^{C}(x_{0},x_{1})-J_{m}^{C}(x_{2},x_{3})=0$ and  $\bar{J}_{m}(x_{0},x_{1},x_{2},x_{3})=\bar{J}_{m}^{C}(x_{0},x_{1})-\bar{J}_{m}^{C}(x_{2},x_{3})=0$.  We find $3q(q-1)$ more nodes in these mirror pairs than in the Chmutov hypersurfaces.  Also the polynomials $J_{m}^{C}(x,y)$ can be of interest in the study of singularities of type $A_{j}, j>1$.

 \begin{figure}[h]
 \includegraphics[width=40pc]{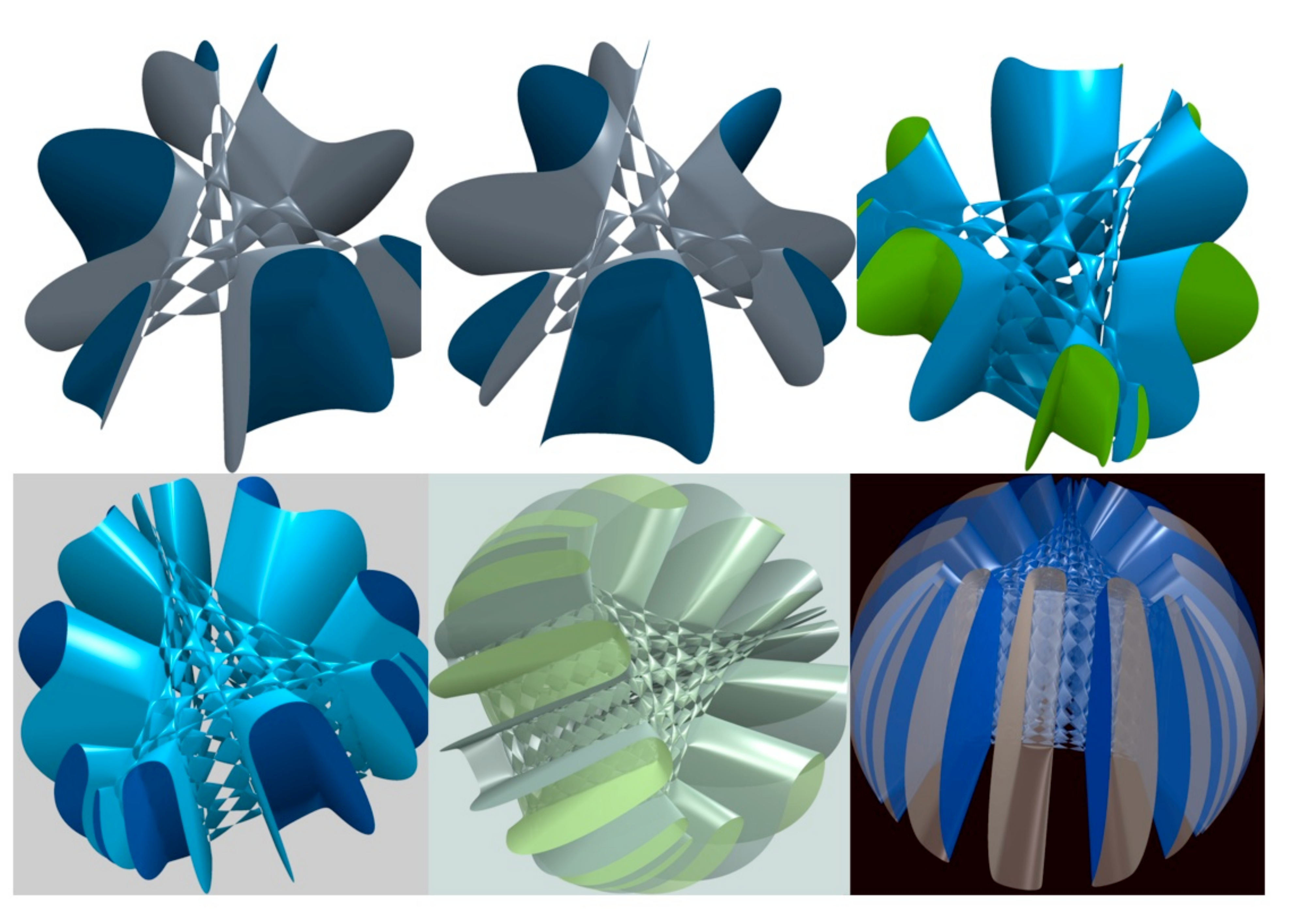}
\caption{\label{label}Two (mirror) sextics with 59 real nodes, a nonic with $\mu_{\bf{R}}=220$, a dodecic with $\mu_{\bf{R}}=581$, a pentadecic surface with $\mu_{\bf{R}}=1162$ and a degree 18 surface with $\mu_{\bf{R}}=2105$.}
\end{figure}

\end{document}